# Characterization and Verification Environment for the RD53A Pixel Readout Chip in 65 nm CMOS

**M. Vogt**[*], **H. Krüger, T. Hemperek, J. Janssen, D.-L. Pohl, M. Daas**
*Universität Bonn*
*Physikalisches Institut*
*Nußallee 12, 53115 Bonn, Germany*
*E-mail:* vogt@physik.uni-bonn.de, krueger@physik.uni-bonn.de,
hemperek@physik.uni-bonn.de, janssen@physik.uni-bonn.de,
pohl@physik.uni-bonn.de, daas@physik.uni-bonn.de

The RD53 collaboration is currently designing a large scale prototype pixel readout chip in 65 nm CMOS technology for the phase 2 upgrades at the HL-LHC. The RD53A chip will be available by the end of the year 2017 and will be extensively tested to confirm if the circuit and the architecture make a solid foundation for the final pixel readout chips for the experiments at the HL-LHC. A test and data acquisition system for the RD53A chip is currently under development to perform single-chip and multi-chip module measurements. In addition, the verification of the RD53A design is performed in a dedicated simulation environment. The concept and the implementation of the test and data acquisition system and the simulation environment, which are based on a modular data acquisition and system testing framework, are presented in this work.

---

[*]Corresponding author.

# 1. Introduction

The inner pixel detector layers play an especially crucial role in precisely identifying the primary vertex of the rare physics events from a large number of particle interactions within the detector. The phase 2 upgrade of the LHC will substantially increase the instantaneous luminosity. In order to meet the experimental challenges of the increased luminosity, the experiments at the LHC will need to address the aging of the present detectors and need to improve their ability to cope with the high-density particle environment and large amounts of radiation. The significant increase of the pile-up density requires novel pixel readout chips with highly complex digital architectures, which deliver hit information at drastically increased data rates and unprecedented radiation tolerance, especially close to the interaction point. The RD53 collaboration was formed to approach these challenges by designing a prototype pixel readout chip in a 65 nm CMOS technology, which is suitable for the innermost layers of the pixel detector in the ATLAS and CMS experiments [1].

The CMOS design of the RD53A chip contains analog and digital circuits [2]. The custom designed analog circuit consists of an Analog Front End (AFE) and bump pads for electrical and mechanical connection to a flip-chip bonded sensor. Three different AFE designs are implemented in the prototype to evaluate the performance of each design. The digital circuit is synthesized from a behavioral description written in SystemVerilog. During the automated logic synthesis, the desired behavior is processed into a CMOS layout, which contains approximately 250 million transistors.

# 2. The RD53A Data Acquisition System

In order to evaluate the design and to measure and characterize the prototype chips, the test and data acquisition system and test bench environment aim to cover the following topics:
- Design verification of the RD53A chip
- Characterization and efficiency measurements in laboratory and test-beam experiments
- Testing of the on-chip power regulators and serially powered multi-chip modules
- Large-scale wafer-level tests on automated probe stations

## 2.1 Hardware

The data acquisition (DAQ) hardware (figure 1) consists of a custom designed base board and a commercially available FPGA module, which is based on a Xilinx Kintex 7 FPGA. Several aspects of the existing MMC3 base board design [3] must be adapted to the needs of the RD53A chip in order to cope with the data rates of the multi-lane serial link with data rates of 1.28 Gbit/s per lane. The RD53A design supports up to four lanes and uses the Aurora 64b66b protocol for serial communication. Therefore, the design of the base board must provide support for several multi-gigabit transceivers channels, additional connectors for the high-speed data links, and an SFP+ port for 10 Gbit/s Ethernet communication required for the transmission of the data to the DAQ computer.

A custom designed PCB (figure 1) carries the RD53A chip in order to interface with the DAQ system. The RD53A chip is glued onto the PCB and wire bonds provide electrical signals and power. Different powering schemes are included in the PCB design, specifically providing support for the dedicated on-chip shunt-LDO power regulators. The PCB is equipped with connectors for the high-speed data links, debugging, and monitoring purposes.

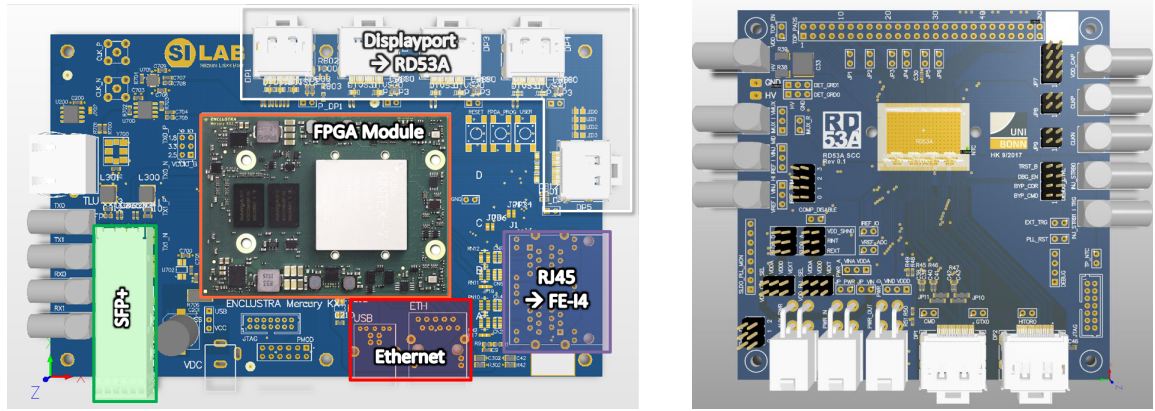

(a) The base board and the FPGA board. Orange: Enclustra Mercury+ KX2 FPGA board. White: DisplayPort connectors for high-speed interconnection to the RD53A chip. Purple: RJ45 connectors for interconnection to the FE-I4 chip. Red: USB 3.0 and 1G Ethernet interfaces. Green: 10G SFP+ port for the connectivity to the DAQ computer.

(b) The Single Chip Card PCB for the RD53A chip. Top side: Digital debugging and monitoring signals. Left and right sides: Analog monitoring signals. Bottom side: Molex power connectors and DisplayPort connectors.

Figure 1: The BDAQ53 hardware.

## 2.2 Software

The BDAQ53 software and firmware [4] is written in Python and is based on the open source data acquisition and system testing framework Basil [5]. The Basil framework provides numerous general purpose FPGA firmware modules (e.g., SPI, FIFO, I/O controllers) and the corresponding drivers. Additionally, lab appliances (e.g., signal generators, digital meters) are supported and can be controlled remotely.
The BDAQ53 software contains specific methods to read and write to the RD53A configuration registers and routines to process and store the received data. Specific calibration and characterization routines for the RD53A chip are currently being developed.

## 2.3 Firmware

The firmware uses Verilog modules, provided by the Basil framework. Several common firmware modules like the $I^2C$ interface or the block memory are instantiated from this resource. The command encoder and the Aurora receiver, which are specific to RD53A chip, have been developed from scratch (figure 2). At the same time, the corresponding counterparts in the RD53A digital design were developed and tested against the DAQ firmware within a simulation environment (section 3). Every module is connected to the proprietary Basil control bus, while the high bandwidth data path utilizes the ARM AXI4-Stream protocol.

In order to facilitate the firmware design and reuse, when moving from simulation to the physical test system, the project is organized into two separate components. The core component contains the functional logic, e.g., state machines, block memory and command encoder, while the interface component contains the modules for Ethernet communication, I/O drivers and PLLs. During simulation, the interface component is replaced by a test bench in which the core component is instantiated (figure 3).

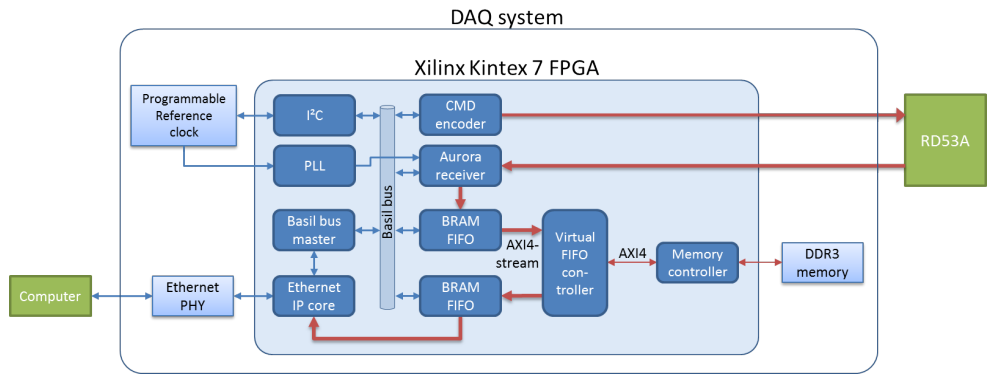

Figure 2: Overview of the FPGA firmware design for the final DAQ software. Dark blue: Firmware elements implemented inside the FPGA. Light blue: Relevant components of the DAQ hardware. Green: External devices.

## 3. Integrated simulation environment

The development of the DAQ software and firmware and the RD53A design are carried out by two different development teams. The DAQ firmware and RD53A digital design are written in a Hardware Description Language (HDL) and have separate simulation environments and test benches for behavioral verification. While the DAQ firmware is synthesized into a bitstream file to configure the DAQ FPGA, the RD53A digital design is synthesized into CMOS logic for the production of the RD53A chip. The interface and protocol specifications ensure compatibility between the DAQ and the RD53A chip.

An essential goal was to combine these two code repositories, to use them within an integrated simulation environment, and to utilize the DAQ software and firmware throughout the RD53A design verification process. This was achieved by integrating the cosimulation test bench environment Cocotb [6] into Basil and by including the register-transfer level (RTL) model of the RD53A chip into the simulation (figure 3). Cocotb is compatible with both open source and proprietary commercial simulators.

The test routines for Cocotb are written in Python, enabling access to a variety of software modules for data processing and analysis. The test routines communicate with the DAQ firmware by using the driver functions of the test system software. During simulation, the Ethernet firmware module is replaced by a simulation interface driver, which is part of the Basil framework. No further firmware modifications are required. Since the test system software is interface-independent, the same functions for register access of the RD53A chip and data processing can be used during simulation as well as for measurements.

Direct stimulation of RD53A pixels are possible, e.g., by injecting pulses into the digital pixel matrix and thus simulating a detected charge in the corresponding sensor pixel.

This approach provides valuable feedback to the design team and allows for a more efficient DAQ system development because the DAQ software and firmware can be evaluated together with the behavioral model of the chip prior to the submission to the semiconductor foundry. This simulation environment can also be used to confirm the results of the already developed SystemVerilog-UVM based verification environment [7] and vice versa.

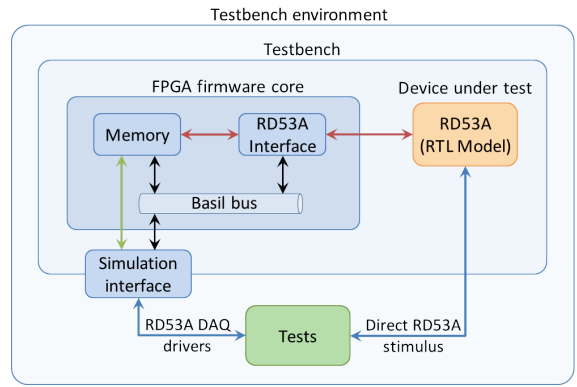

Figure 3: The cosimulation test bench environment and overview of the modified FPGA firmware that allows for communication with the test bench procedures via the simulation interface driver.

## 4. Summary and Outlook

The RD53A simulation environment and several test routines have been developed to verify the functionality of the RD53A digital design. The interconnection between the DAQ system and RD53A chip has been tested in the simulation environment and the operability of the Aurora 64b66b connection in single-, two-, and four-lane configuration have been verified. Several Aurora-related communication issues were discovered in the behavioral description, which could be fixed before tape-out. By stimulating the RD53A digital pixel matrix, it was shown that hit signals are processed and event data frames are generated correctly.

The RD53A chip was submitted for production in August 2017 and the samples are expected to be shipped by the end of November 2017. In the meantime, the further development of the test and data acquisition software is ongoing, and the software is tested using the RD53A simulation environment. As soon as the RD53A chips become available, characterization measurements will be performed with the DAQ software.

### Acknowledgments

This project has received funding from the European Union's Horizon 2020 Research and Innovation programme under Grant Agreement no. 654168.